\documentclass[a4paper]{jpconf}
\usepackage{graphicx}
\usepackage{subfigure}
\begin{document}
\title{First-order transition in {\it XY} model with higher-order interactions}

\author{Milan \v{Z}ukovi\v{c}}

\address{Institute of Physics, Faculty of Science, P.J. \v{S}af\'arik University, Park Angelinum 9, 041 54 Ko\v{s}ice, Slovakia}

\ead{milan.zukovic@upjs.sk}

\begin{abstract}
The effect of inclusion of higher-order interactions in the {\it XY} model on critical properties is studied by Monte Carlo simulations. It is found that an increasing number of the higher-order terms in the Hamiltonian modifies the shape of the potential, which beyond a certain value leads to the change of the nature of the transition from continuous to first order. The evidence for the first-order transition is provided in the form of the finite-size scaling and the energy histogram analysis. A rough phase diagram is presented as a function of the number of the higher-order interaction terms.
\end{abstract}

\section{Introduction}

The {\it XY} spin model is well known to undergo a topological Berezinskii-Kosterlitz-Thouless (BKT) phase transition to a low-temperature quasi-long-range-order (QLRO) phase characterized by an algebraically decaying correlation function~\cite{bere71,kost73}. The standard {\it XY} model only includes bilinear pairwise couplings between neighboring spins. Nevertheless, it has been shown that higher-degree pair interactions between magnetic ions may be comparable or even stronger than the bilinear ones (see, e.g., Refs.~\cite{chiu75,fedd79}, and references therein) and thus can play an important role in the system's behavior. They turned out useful in modeling of certain systems, such as liquid crystals~\cite{lee85,geng09}, superfluid A phase of $^3{\rm He}$~\cite{kors85}, and high-temperature cuprate superconductors~\cite{hlub08}. These findings resulted in an increased interest in the study of the model with higher-order interactions. The model involving terms up to the $p$-th order can be described by the Hamiltonian

\begin{equation}
\label{Hamiltonian1}
{\mathcal H}=-\sum_{k=1}^{p}J_{k}\sum_{\langle i,j \rangle}({\bf S_i}{\bf S_j})^k,
\end{equation}
where $J_{k}$ are the $k$-th order exchange coupling constants, ${\bf S_i}=(S_i^x,S_i^y)$ is a two-dimensional unit vector at site $i$, and $\langle i,j \rangle$ denotes the sum over nearest-neighbor spins. In the present study we will assume that all the couplings are of equal strength and normalized, i.e., $J_{k} \equiv J=1/p$. Then, the Hamiltonian can be rewritten in the form 
\begin{equation}
\label{Hamiltonian2}
{\mathcal H}=-J\sum_{\langle i,j \rangle}\frac{\cos\theta_{i,j}(1-\cos^{p}\theta_{i,j})}{1-\cos\theta_{i,j}},
\end{equation}
where $\theta_{i,j}=\theta_{i}-\theta_{j}$ is an angle between the nearest-neighbor spins.

The most studied was the bilinear-biquadratic model, i.e., the above model for $p=2$~\cite{lee85,kors85,carp89,shi11,hubs13,qi13}. Inclusion of a sufficiently large biquadratic coupling has been shown to lead to the separation of the quadrupole phase at higher temperatures from the dipole phase at lower temperatures. Nevertheless, the character of the phase transition to the paramagnetic phase, either from the dipole or quadrupole phases, was confirmed to belong to the BKT universality class, while the dipole-quadrupole phase transition was of the two-dimensional Ising universality class.

Motivated by orientational transitions in liquid crystals, the inclusion of the higher-order terms has also been taken in the form of the the $k$-th order Legendre polynomial of the bilinear term~\cite{fari05,berc05,fari10}. Although in models with $O(n)$ symmetry for $n \geq 2$ a crossover to the first order transition is expected for large enough values of $k$~\cite{ente02,ente05}, for the case of $O(2)$ and the studied values of $k=2$ and $4$ no such crossover has been confirmed.

Domany {\it et al.}~\cite{doma84} have introduced a modified {\it XY} model that enabled tuning its critical properties through the parameter $p^2$ between the standard {\it XY} model belonging to the BKT universality and the $q$-state Potts model. The latter is known to show a first-order transition for large $q$ and a crossover to the first-order was also confirmed for large $p^2$ in this modified {\it XY} model~\cite{himb84,blot02,sinh10a,sinh10b}.

In this study we generalize the {\it XY} model by including up to the $p$-th order isotropic pairwise interactions to the bilinear term. We show that, in spite of belonging to the same universality class as the standard {\it XY} model, the generalized model changes the character of the transition to the paramagnetic phase to the first-order one for sufficiently large $p$.

\section{Method}
We use Monte Carlo (MC) simulations for spin systems on a square lattice of a size $L \times L$, with $L=24-120$ and periodic boundary conditions. The spin update follows the Metropolis algorithm with the following parameters: the initial $4 \times 10^4$ MC sweeps are discarded for thermalization and the following $2\times 10^5$ MC sweeps are used for the calculation of thermal averages. Temperature dependencies of various measured quantities are obtained by starting the simulations at sufficiently high temperatures $T$ (measured in units $J/k_B$, where $k_B$ is the Boltzmann constant), corresponding to the paramagnetic phase. Then the temperature is gradually decreased by $\Delta T=0.025$ and the simulation at $T-\Delta T$ is initialized using the last configuration obtained at $T$.  

In the vicinity of the transition to the paramagnetic phase we perform a finite-size scaling (FSS) analysis by using the reweighting techniques~\cite{ferr88,ferr89}, to convincingly identify the order of the transition. For the reweighting we use much longer simulation runs, including up to $10^7$ MC sweeps.

Two basic quantities are measured: the internal energy per spin 
\begin{equation}
e=\langle {\mathcal H} \rangle/L^2
\label{e}
\end{equation}
and the magnetization per spin
\begin{equation}
m=\langle M \rangle/L^2=\left\langle\Big|\sum_{j=1}^{L^2}\exp(i\theta_j)\Big|\right\rangle/L^2.
\label{m}
\end{equation}
To determine the transition order we also evaluate the magnetic susceptibility
\begin{equation}
\label{chi}\chi = \frac{\langle M^{2} \rangle - \langle M \rangle^{2}}{L^2T}, 
\end{equation}
and the fourth-order magnetic cumulant
\begin{equation}
\label{U}U = 1-\frac{\langle M^{4}\rangle}{3\langle M^{2}\rangle^{2}}.
\end{equation} 

If the transition is of first order, then at the transition point the internal energy $e$ and the magnetization $m$ will show a discontinuous behavior, the thermodynamic functions like the susceptibility $\chi$ will scale with volume, i.e., $\chi(L) \propto L^{2}$, and the fourth-order magnetic cumulant $U$ will show an abrupt dip to negative values~\cite{tsai98}.

\section{Results}
\begin{figure}[t]
\begin{center}
\begin{minipage}{36.4pc}
\subfigure{\includegraphics[width=18.2pc]{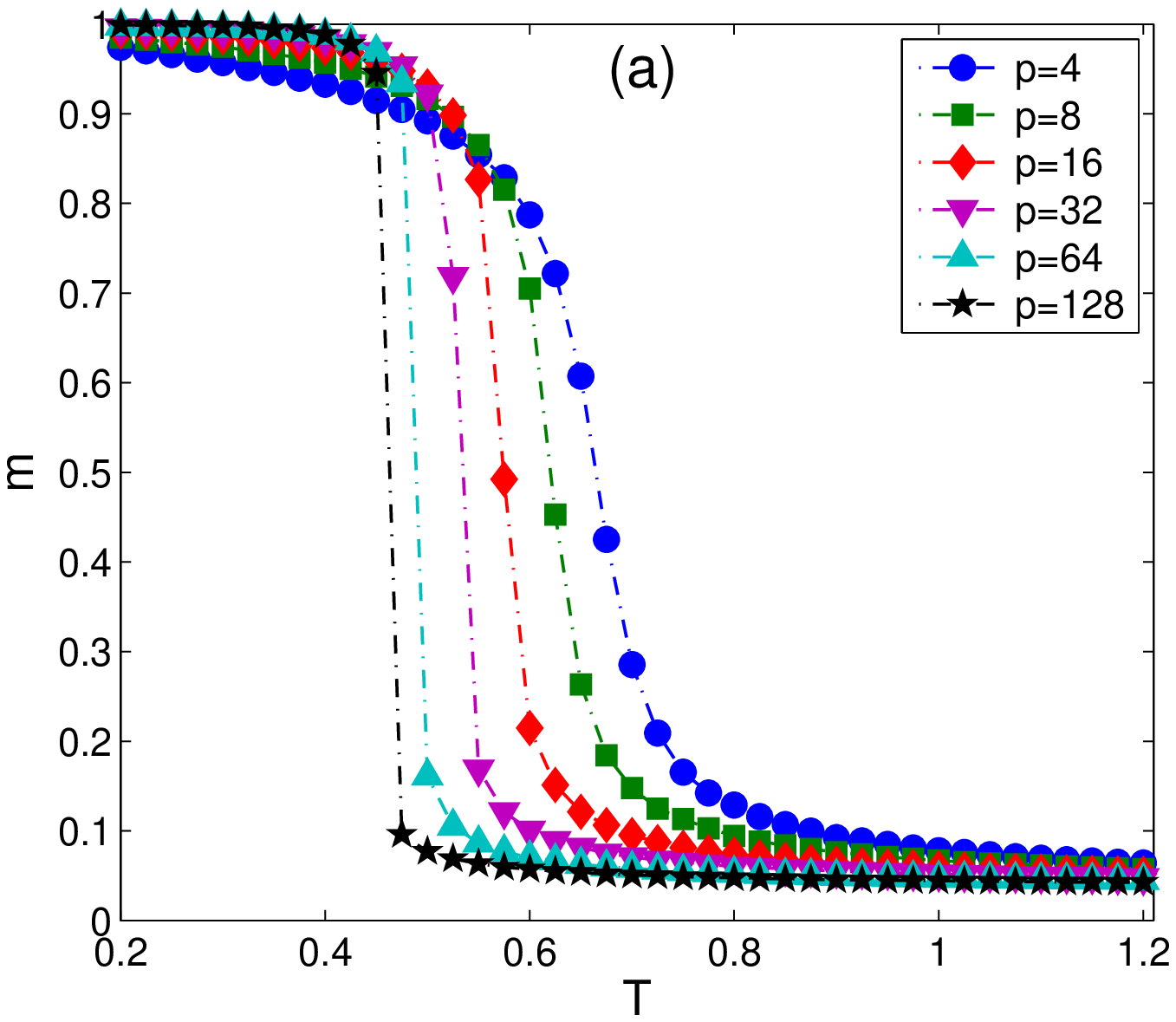}\label{fig:m-T}}
\subfigure{\includegraphics[width=18.2pc]{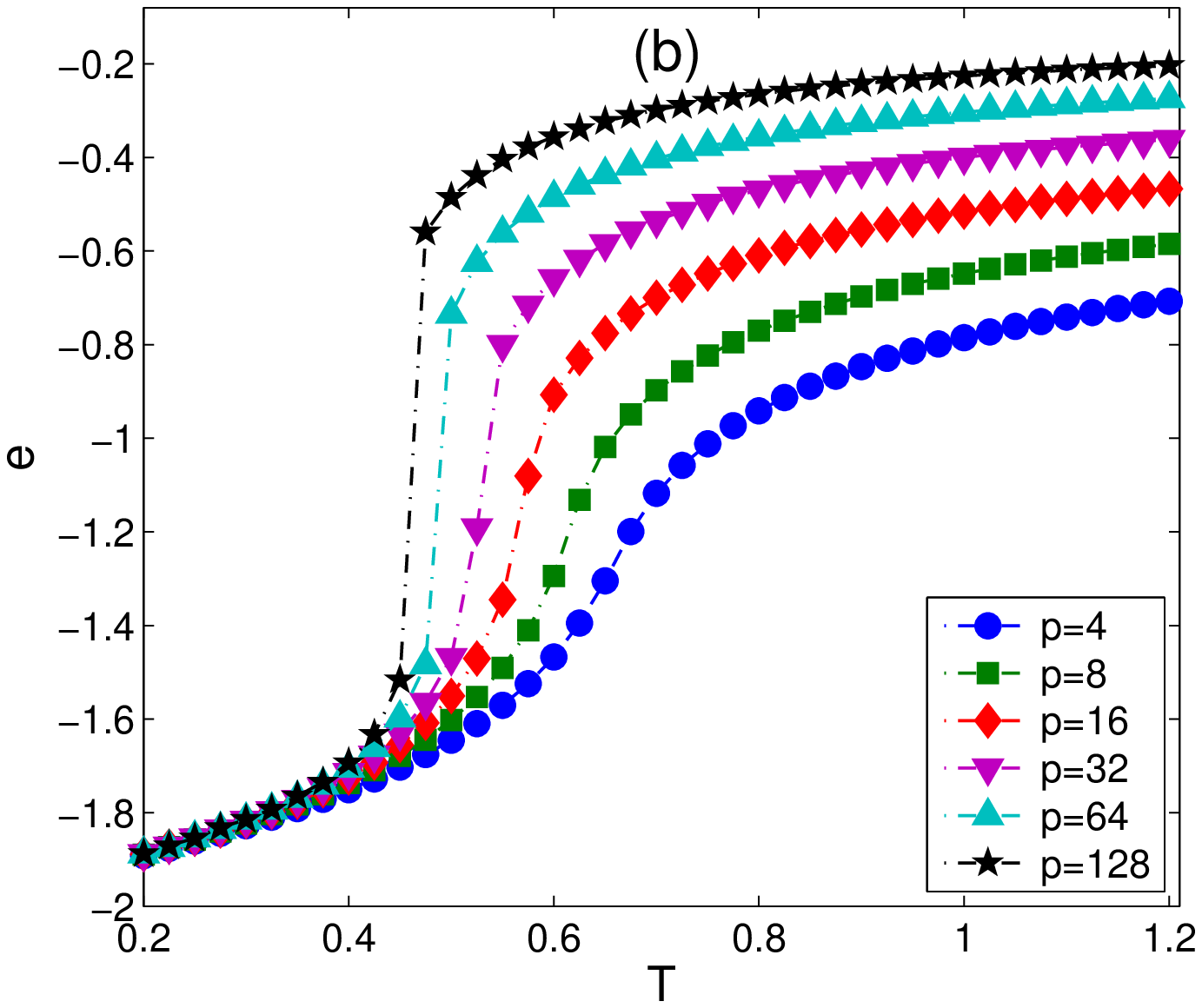}\label{fig:e-T}}
\caption{\label{fig:x-T}Temperature dependence of (a) the magnetization and (b) the internal energy, for $L=24$ and several values of $p$.}
\end{minipage}
\end{center}
\end{figure}

Temperature dependencies of the basic quantities, i.e., the internal energy and the magnetization, are presented in Figs.~\ref{fig:m-T} and~\ref{fig:e-T}, for $L=24$ and the increasing number of the higher-order interactions $p=2,8,16,32,64$ and $128$. One can observe a transition between the high-temperature paramagnetic phase, characterized by $m \approx 0$~\footnote{$m=0$ is only expected in the thermodynamic limit of $L \to \infty$.}, and the low-temperature QLRO phase with $m>0$. As $p$ increases, close to the transition point the behavior of both quantities changes from a slow variation (similar to the {\it XY} model) to more abrupt changes. For $p=64$ and $128$, the variation appears clearly discontinuous, as in the case of first-order transitions.

\begin{figure}[t]
\begin{center}
\begin{minipage}{18.2pc}
\includegraphics[width=18.2pc]{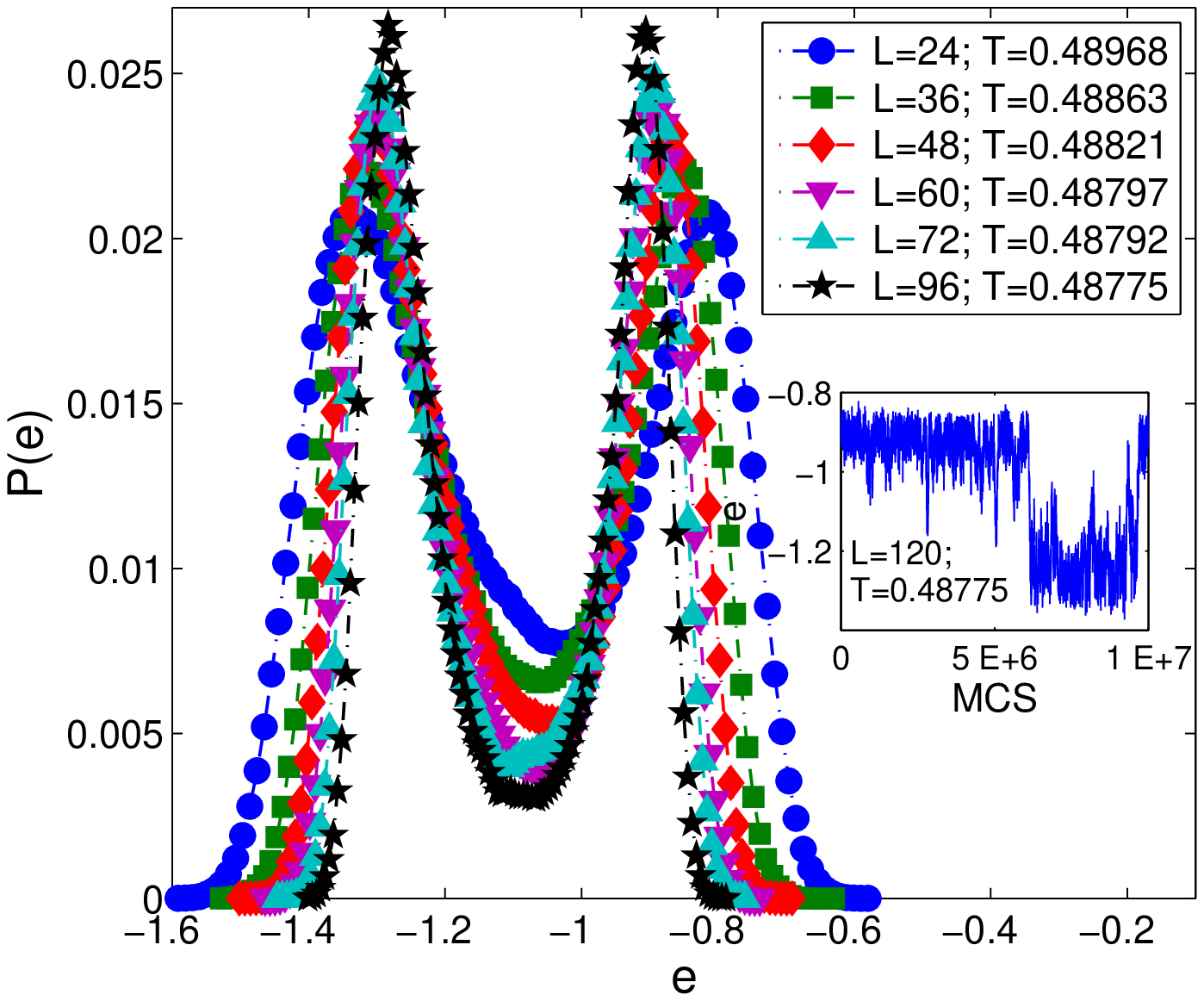}
\caption{\label{fig:hist} Energy histograms at the pseudo-transition points, for $p=64$ and various $L$. The inset demonstrates enormous tunneling times between the coexisting states, for $L=120$.}
\end{minipage}\hspace{1.5pc}%
\begin{minipage}{18.2pc}
\includegraphics[width=18.2pc]{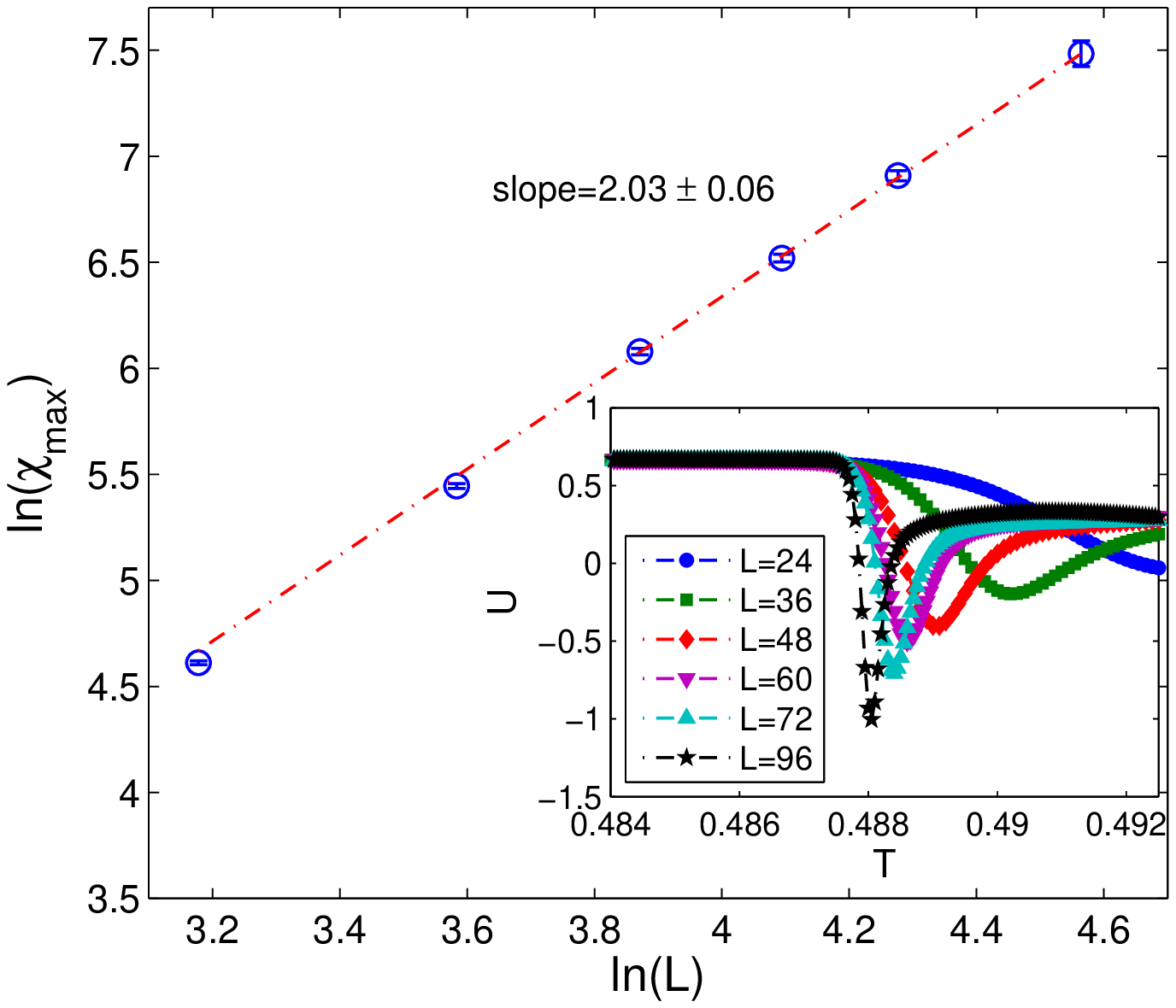}
\caption{\label{fig:FSS} FSS of the magnetic susceptibility at the transition point, for $p=64$ and various $L$. The inset shows the temperature dependence of the fourth-order magnetic cumulant.}
\end{minipage} 
\end{center}
\end{figure}

In order to make certain that the apparent discontinuity in the magnetization and energy is not an artifact of the finite system size, we additionally performed very long runs ($10^7$ MC sweeps) close to the transition points and calculated energy histograms. As evidenced from Fig.~\ref{fig:hist}, already for $p=64$ the energy distributions shows a bimodal character that is characteristic for a discontinuous first-order transition. The histograms for different sizes $L$ are reweighted to the temperatures at which both peaks are of equal height. These temperatures correspond to the pseudo-transition temperatures $T_c(L)$ and the true transition temperature can be obtained by extrapolation to the thermodynamic limit, i.e., for $L \to \infty$. The scaling properties of the energy distribution are in accordance with what we expect from the first-order transition. Namely, as $L$ increases the heights of the peaks increase at the cost of the dip (barrier) between them. The latter tends to zero and the distance between the peaks approaches a finite value, corresponding to the latent heat released at the discontinuous transition. We note that the histogram for the largest considered lattice size $L=120$ is not presented due to enormous tunneling times (see the inset), which, for the considered simulation times, prevent a reliable histogram estimation.

To obtain an additional evidence for the first-order transition scenario, we also perform FSS analysis and inspect the behavior of the fourth-order magnetic cumulant. As presented in Fig.~\ref{fig:FSS}, the magnetic susceptibility is confirmed to scale with volume, as it should be in the case of a first-order transition. The inset of the same figure shows the temperature variation of the fourth-order magnetic cumulant. An abrupt descent to negative values at the transition point corroborates the above findings and provides further support for the first-order transition scenario. 

As can be predicted from Fig.~\ref{fig:x-T}, for larger $p$ the first-order transition features become even more pronounced. The energy barrier between the two coexisting phases becomes deeper and the gap between the peak's positions becomes wider (not shown). On the other hand, the case of $p=1$ corresponds to the standard {\it XY} model showing a continuous phase transition. Therefore, we expect the existence of a value of $1<p_c<64$ that would correspond to the tricritical point at which the transition changes its nature between the first and second order. A more precise location of $p_c$ may not be trivial due to possible problems with a reliable determination of the transition order in the crossover region (near the expected tricritical point) and we leave this task for further investigation. 

\begin{figure}[t]
\centering
\begin{minipage}{18.2pc}
\includegraphics[width=18.2pc]{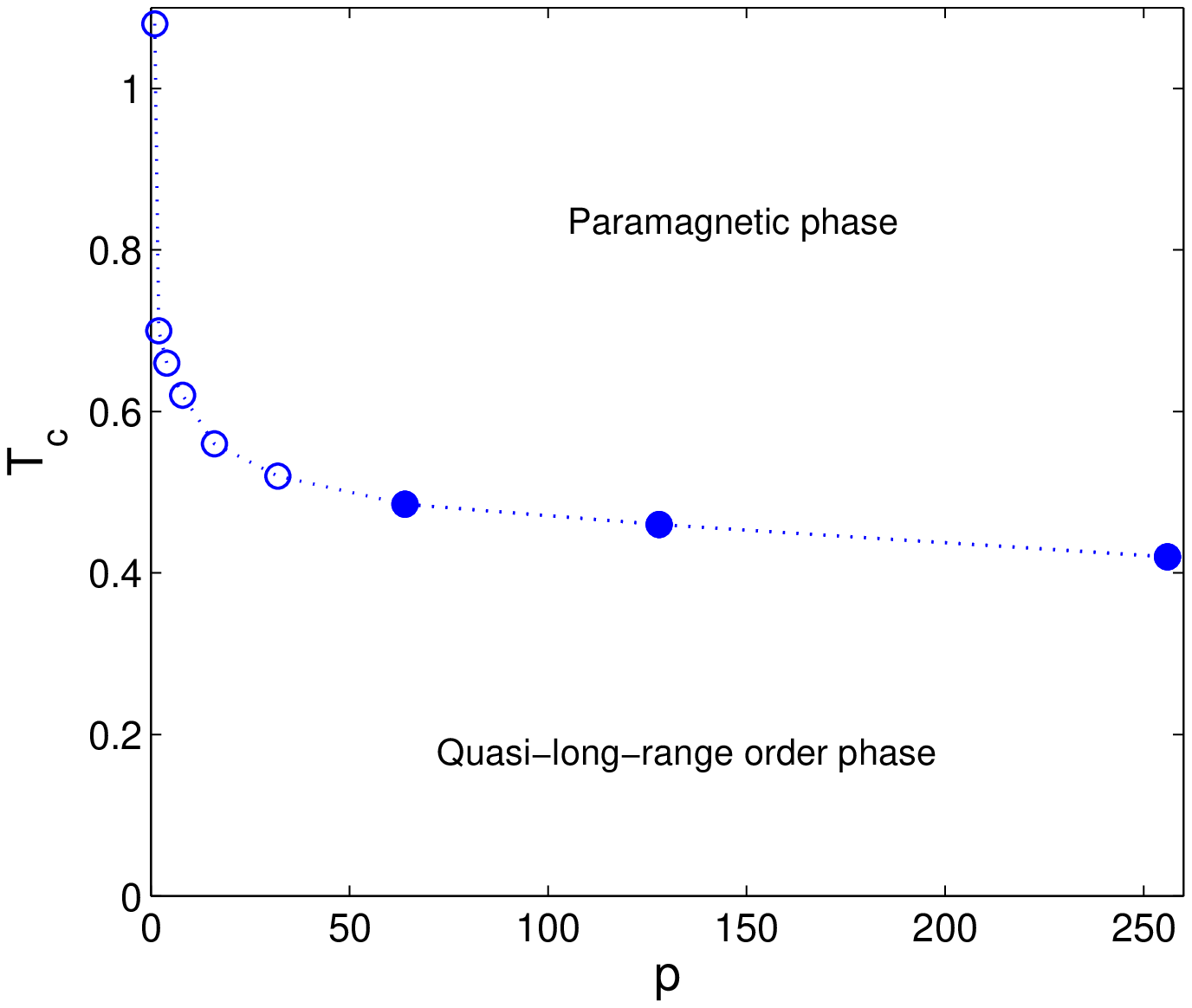}
\caption{\label{fig:PD} Phase diagram. The filled symbols show first-order transition points and the dotted line is a guide to the eye.}
\end{minipage}\hspace{1.5pc}%
\begin{minipage}{18.2pc}
\includegraphics[width=18.2pc]{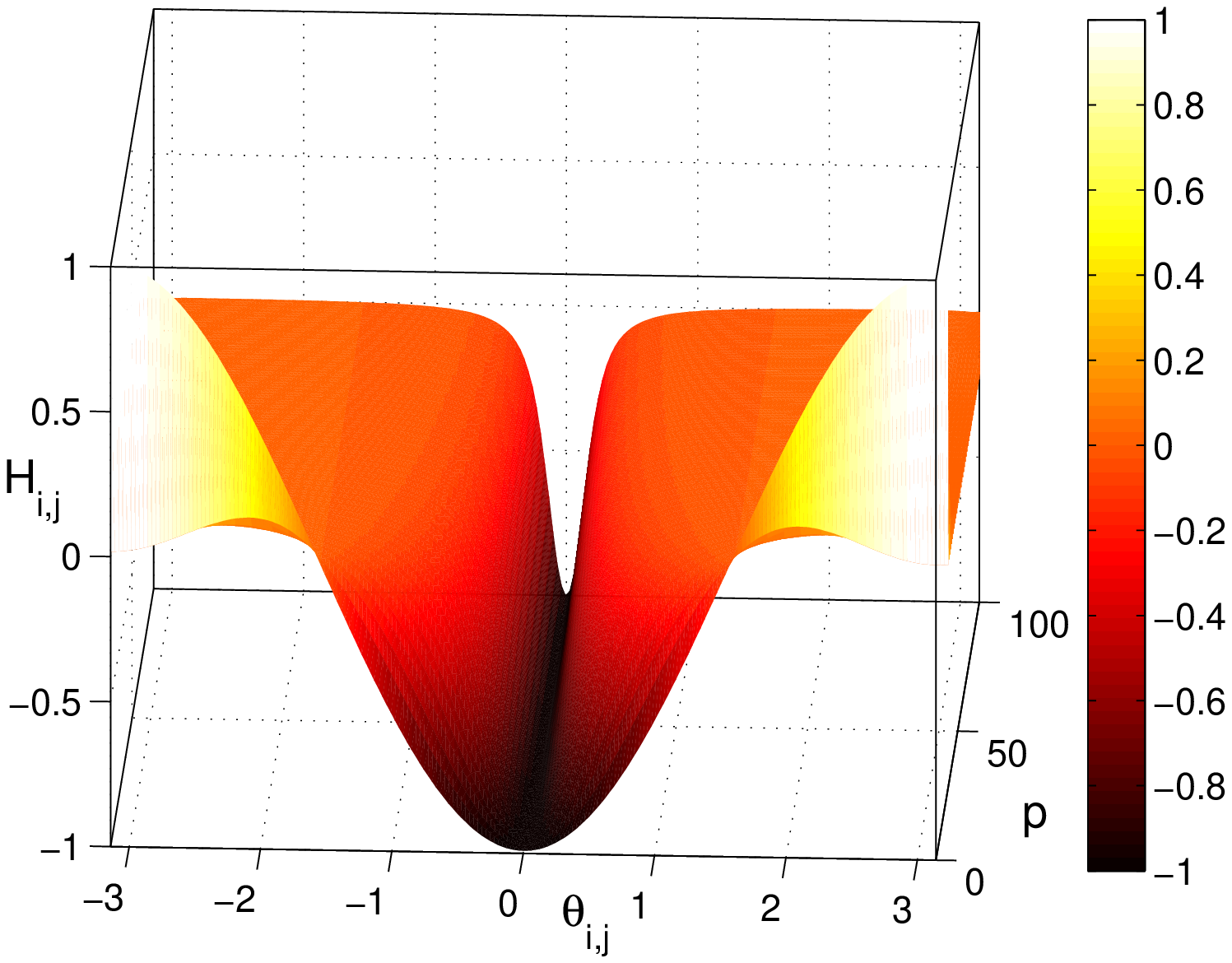}
\caption{\label{fig:en_well} Potential well as a function of the neighboring spin angle $\theta_{i,j}$ and the number of higher-order interaction terms $p$.}
\end{minipage}
\end{figure}

An approximate phase diagram in $T-p$ parameter plane is depicted in Fig.~\ref{fig:PD}. Rough estimates of (pseudo)transition temperatures are obtained as positions of peaks of the magnetic susceptibility, for a fixed lattice size $L=24$. The filled circles represent the first-order transition points (for $p \geq 64$). We note that these pseudo-transition temperatures slightly overestimate the true thermodynamic limit values, as also evidenced in the {\it XY} model (see, e.g., Ref.~\cite{fari10}). Overall, the increasing $p$ shifts the transition temperature between the paramagnetic and the QLRO phases to lower values and eventually also changes its nature to the first-order one.

\section{Discussion}
Finally, let us briefly discuss the origin of the crossover to the first-order transition regime. We believe that the mechanism leading to the first-order transition is similar to that observed in the nonlinear model introduced by Domany {\it et al.}~\cite{doma84}. It was concluded that the change in the nature of the transition from continuous to first order in that model resulted from an increase of the nonlinearity parameter $p^2$, which changed the shape of the potential. In the present model, the nonlinearity parameter $p$ represents the number of the higher-order interaction terms, nevertheless, its effect on the potential is similar. The dependence of the potential $H_{i,j}=\cos\theta_{i,j}(1-\cos^{p}\theta_{i,j})/[p(1-\cos\theta_{i,j})]$ on the parameter $p$ is depicted in Fig.~\ref{fig:en_well}. Like in the Domany's model, with the increasing $p$ the cosine shape of the potential well for $p=1$ (the standard {\it XY} model) becomes more and more narrow with a width tending to zero. The narrowing of the potential well at low temperatures can be ascribed to insufficient increase in the vertex density followed by their abrupt (discontinuous) proliferation at higher temperatures, leading to a first-order transition. A more detailed study of the vertex formation will be presented elsewhere.
%

\ack
This work was supported by the Scientific Grant Agency of Ministry of Education of Slovak Republic (Grant No. 1/0331/15) and the scientific grants of Slovak Research
and Development Agency provided under contract No.~APVV-0132-11 and No.~APVV-14-0073.

\end{document}